\documentclass[conference]{IEEEtran}
\IEEEoverridecommandlockouts
\usepackage{cite}
\usepackage{amsmath,amssymb,amsfonts}
\usepackage{algorithmic}
\usepackage{graphicx}
\usepackage{textcomp}
\usepackage{xcolor}
\usepackage{subfig}
\usepackage{dblfloatfix}

\def\BibTeX{{\rm B\kern-.05em{\sc i\kern-.025em b}\kern-.08em
    T\kern-.1667em\lower.7ex\hbox{E}\kern-.125emX}}
\begin{document}

\title{Show Me Your Account: Detecting MMORPG Game Bot Leveraging Financial Analysis with LSTM}

\author{\IEEEauthorblockN{Kyung Ho Park}
\IEEEauthorblockA{\textit{Graduate School of Information Security} \\
\textit{Korea University}\\
Seoul, Republic of Korea \\
kyungho96@korea.ac.kr}
\and
\IEEEauthorblockN{Eunjo Lee}
\IEEEauthorblockA{\textit{Data Center} \\
\textit{NCSOFT}\\
Seongnam, Republic of Korea \\
gimmesilver@ncsoft.com}
\and
\IEEEauthorblockN{Huy Kang Kim}
\IEEEauthorblockA{\textit{Graduate School of Information Security} \\
\textit{Korea University}\\
Seoul, Republic of Korea \\
cenda@korea.ac.kr}
}

\maketitle

\begin{abstract}
With the rapid growth of MMORPG market, game bot detection has become an essential task for maintaining stable in-game ecosystem. To classify bots from normal users, detection methods are proposed in both game client and server-side. Among various classification methods, data mining method in server-side captured  unique characteristics of bots efficiently. For features used in data mining, behavioral and social actions of character are analyzed with numerous algorithms. However, bot developers can evade the previous detection methods by changing bot's activities continuously. Eventually, overall maintenance cost increases because the selected features need to be updated along with the change of bot's behavior.

To overcome this limitation, we propose improved bot detection method with financial analysis. As bot's activity absolutely necessitates the change of financial status, analyzing financial fluctuation effectively captures bots as a key feature. We trained and tested model with actual data of Aion, a leading MMORPG in Asia. Leveraging that LSTM efficiently recognizes time-series movement of data, we achieved meaningful detection performance. Further on this model, we expect sustainable bot detection system in the near future.
\end{abstract}


\section{Introduction}
Online game plays a huge role in modern leisure. With 44.6 billion US dollars of market share estimated by 2022, Massively Multiplayer Online Role-Playing Game (MMORPG) takes significant position in global market \cite{AP}. MMORPG is a game genre that users have own characters to play various activities in a virtual world. They combat with monsters, accumulate game assets, even chat or date with others. As various activities exist, in-game ecosystem shows similar pattern as a real world. \cite{castronova2001virtual} Interestingly, people in a real world and MMORPG users pursue a common goal: a wealth. In a real world, diligent workers accumulate wealth from salaries. Similarly, heavy users in a virtual world gather game assets and become wealthy. Wealth in a game makes character stronger, and stronger character gives more fun to player. Reflecting that players want better game assets, some corporate-like entities even created Real Money Trading (RMT), a transaction of game assets with real money \cite{huhh2008simple}. However, money-related issues created illegal activities. Some malicious users developed automated program, called a game bot.

Game bot is an automated program playing game autonomously instead of human users. Without any touch of human, bots automatically move around virtual world. They behave in programmed way, repetitively do patterned actions to collect game asset. Bots normally gather wealth faster than normal players because program does not get tired \cite{lee2016you}. Leveraging this efficiency, corporate-like entities called Gold Farmer Group (GFG) started to operate thousands of bots at the same time. GFGs collect enormous amount of in-game cash or rare items, and sell it to buyers with real money \cite{ahmad2009mining}.

Game companies should detect these bots and provide adequate actions because bots create deprivation of normal users. If some users purchase items with real money and become powerful easily, normal users would depreciate their effort of growing characters. Moreover, in-game economy goes unstable with affluent assets created by bots. GFG especially collects a huge amount of game assets, and creates an inflation and overflow of asset. Inflated assets easily break balance among users and skew the game design. As a result, bots make gaming experience of normal users feel depreciated, leave the game, and eventually create loss to game company.

To counteract malicious activities of bots, academia and industry have developed bot detection methods in two main streams: client-side detection and server-side detection. Client-side method detects bots by implementing challenge-response program or security solution. For challenge-response method, game client asks question that humans can easily solve. As bots are not programmed to answer unexpected questions, characters with wrong response are classified as bots. Security solutions such as GameGuard or Warden are specially developed client-side program for bot detection. However, both methods are inadequate to apply at industry level. Challenge-response methods like CAPTCHA excessively drop user's game experience, as they feel disturbed during the play. Security solutions frequently collide with vaccine programs, create crash of the system \cite{woo2012survey}.

To overcome limits of client-side methods, server-side methods have been proposed. Server-side methods capture unique characteristic of bots leverage data mining. Past works suggested numerous methods with data mining algorithms, but there were several hurdles. First of all, previous methods optimized detection model in a certain game, thus hard to generalize. Models performed well in selected game, but hard to be utilized in other games as well. Moreover, bot developers could neutralize detection methods if they recognize detection thresholds. As bot developers figured out patterns of bot detection, they started to make bots mimicking human user's behavior. For sustainable, generalized, and secured bot detection, it is necessary to develop improved method.

In this work, we propose bot detection method to jump over limitations of previous works. To design sustainable and commonly applicable detection method, we performed financial analysis to each character. Considering that bots eventually accumulate their wealth to specific character for RMT, we analyzed flow of in-game cash and assets. This generates effectiveness in two aspects. First, we can vividly capture bot's behavior as bots cannot repudiate financial patterns. If transactions among bots exist for accumulation, specific financial patterns must exist. Furthermore, financial analysis can be applied in different games as concept of cash or items generally exist. Most MMORPGs operate its own economy, and transactions occur among users. We can easily compare features used in financial analysis, and apply into different games.

Leveraging Recurrent Neural Networks (RNN) as algorithm, we trained model with game play log of Aion, a famous MMORPG developed by NCSOFT. We evaluated our model precisely captures financial pattern of bots, and provide improved detection method with following contributions below:

 1) Sustainability: As financial pattern of character is inevitable record, model effectively detects bots although bot behaves just like normal players or change its activity pattern.

 2) Generality: As model utilizes commonly used features in general MMORPG, game companies can easily apply model into multiple games.
 
 3) Individuality: Only once we train model with accumulated data, model classifies individual character as bot or normal user.

 4) Security: As neural network is a black-box model, bot developers are hard to recognize detection thresholds.

\section{Literature Review}

\begin{table*}[!t]
\caption{Researches on server-side bot detection method}
\label{tab:literatures}
\centering
\resizebox{13cm}{!}{
\begin{tabular}{|l|l|l|c|}

\hline
Category & Data type & Modeling & Reference\\

\hline
Sufficient Condition & Behavioral Action & Sequence Pattern & \cite{lee2015game} \\
& & Self-similarity of Action & \cite{lee2016you} \\
& & Action Frequency & \cite{thawonmas2008detection} \\
& Social Action & Chat Log Analysis & \cite{kang2012chatting}\\
& & Party Play Log Analysis & \cite{kang2013online}\\
\hline
Necessary Condition & Transaction & Network Analysis & \cite{song2015game}\\
& Coordinates where asset & DBSCAN & \cite{lee2018no}\\
& increase or decrease & & \\
\hline

\end{tabular}
}
\end{table*}

Among previous works, server-side method analyzes game play data to capture unique pattern of bots. We categorized features used in bot detection as Table~\ref{tab:literatures}, by dividing into two streams: sufficient condition and necessary condition. The sufficient conditions are set of features that bot's activity probably creates, but not absolutely accompanies. As the sufficient conditions include distinctive pattern of bots, we can identify bots during certain period of time. However, analysis of the sufficient conditions require repetitive updates as these conditions are not consistent. If bot developers modify bot's behavior, model also necessitates update to capture changed pattern. On the other hand, the necessary conditions are features must happen as a consequence of bot's behavior. As bots are designed to accumulate wealth efficiently, specific actions such as transaction among bots absolutely happen. Although bot developers change bot's behavior, pattern of wealth accumulation itself still exists. Thus, we can detect bots leveraging necessary conditions in consistent way. 

Detection method with sufficient conditions analyzes two data types: behavioral action and social action. Behavioral actions describe how character performs physical activities such as moving, normal attack, or using skills. Lee \textit{et al.} analyzed the full action sequence of users on big data analysis platform. They set specific behavior sequences and applied simple scoring algorithm and Naive Bayesian algorithm to classify bots from users \cite{lee2015game}. In another method, Lee \textit{et al.} captured similarity of character behavior to classify game bots. They divided sequence of behavior with time window, and embedded as a feature vector. They showed bots have similar behavior pattern during play time by applying logistic regression algorithm with self-similarity of characters \cite{lee2016you}.

Not only behavioral characteristics, social actions also illustrate distinguishable patterns. Thousands of users communicate and socialize with others. They chat, form a party to complete quests, or create guild for sense of belonging. Leveraging chat logs among users, Kang \textit{et al.} derived lexical, syntactic, semantic features from chatting contents using text mining methods. As bots have similar pattern of chatting to evade detection rules, analyzing text features with machine learning algorithms showed such high performance \cite{kang2012chatting}. Kang \textit{et al.} also analyzed party play log for game bot classification. They focused that the party play in MMORPG requires strong interaction between game players in a short time, which creates different party play pattern between bots and normal users. By inspection that game parties composed of bots play in distinctive way, they established thresholds for bot detection statistical algorithms \cite{kang2013online}.

Above methods with sufficient conditions showed significant performance, but accompanied limitation of sustainability. If bot developers change pattern of game bots, detection methods are easily avoidable. In this circumstance, we should change detection rule or re-train model repetitively. Some bots in these days started to behave and socialize like normal users. Intelligent bots are programmed to generate plausible chats, or do actions like normal users. Furthermore, if game company updates in-game ecosystem, it might blur existing detection rules. As bots change their behavior or actions following update, detection methods also require update of its threshold. Therefore, we acknowledged necessity of sustainable detection method using different features.

To overcome this limitation, some researches suggested detection method with necessary conditions. Financial features such as character's asset level and transaction are actively used. Song and Kim captured geographical tendency that location of transaction among bots show similar pattern. They analyzed specific coordinates of financial transaction within a map, and applied Density-Based Spatial Clustering of Applications with Noise (DBSCAN) algorithm to identify bots from normal users \cite{song2015game}. Lee \textit{et al.} build a topological network of all transactions in a virtual world, and figured out specific patterns of transactions among bots. Especially in GFG, they illustrated bots have different roles for efficient asset gathering, and proposed structure of network for clear understanding of bot ecosystem \cite{lee2018no}.

Analysis of necessary conditions showed financial features are meaningful to identify bots from normal users. However, bot developers still can hedge detection method by changing transaction medium. For instance, if bots send cash or items through mailing, it does not leave location coordinates of transaction. Topological network analysis revealed macroscopic understanding about transactions among bots, but hard to capture individual bots rapidly. Building a network requires heavy resources to analyze whole structure. Whole network also necessitates repetitive update following change of transaction patterns among bots.

To improve previous methods, we present a bot detection method leveraging financial analysis. In pursuit of sustainable detection method, we analyzed financial status of character, necessary conditions of bot's activity. Level of cash or number of items a character owns are examples of financial status. We utilized status data rather than other features, as status itself cannot be modified. Derived features such as transaction coordinates can be hedged or easily changed. But status itself is a consequence of transaction, thus inevitable by bot developers. To scrutinize individual bot's financial data, we employed Long Short-Term Memory neural networks (LSTM) as algorithm. As neural network requires less resource rather than topological network, it enables economic model establishment process. From following sections, we propose bot detection model validated with actual MMORPG play data.

\section{Proposed Methodology}

\subsection{Data Collection}
We collected the game play dataset from Aion, one of the most famous MMORPGs in the world. The dataset is accumulated during the first week of May, 2010. Through collection process, we complied the End User License Agreement (EULA) and related laws under consent of Aion users. Anonymous data are confidentially collected, and utilized only for analysis of this work.

\subsubsection{Feature selection}
To filter features with financial status, we dropped unnecessary information through feature selection. In log data, there exists status log which periodically shows character's information. Among various status features described in Table~\ref{tab:features}, we extracted financial information, which illustrates financial status of character.

\begin{table}
\caption{Feature types in Aion status log}\label{tab:features}
\centering

\begin{tabular}{|l|l|}
\hline
Type & Detailed features\\
\hline
Identification & Character number, Account number\\

\hline
Location & Location coordinates, Map number\\

\hline
Playing Information & Health Point, Magical Point, Experience point\\

\hline
Social Information & Party identification number,\\
& Alliance identification number\\

\hline
Financial Information & Cash status, Item status, Inventory status\\

\hline
\end{tabular}%
\end{table}

\subsubsection{Ground-truth confirmation}
Confirming ground-truth is an essential process of bot detection. We labeled data as bot and normal user following judgement of game company, NCSOFT. Company operates human inspectors who observe doubtful characters by hand. To evade mistakes and bias of human observers, company carefully labels doubtful character only when multiple observers decided in a similar way. Company also collects reasons of decision for comprehensive understanding. If blocked character was actually a normal user, character label is updated as normal user again. As our ground-truth is constructed through these sophisticated labelling process, we regard solid ground-truth for detection model is confirmed.

\begin{table}
\caption{Rules for identifying non-influential features}\label{tab:rules}
\centering

\begin{tabular}{|c|l|l|}
\hline
No & Rule & Description\\

\hline
1 & Feature indifference & A value of a feature is indifferent at\\
& & bot and normal user\\

\hline
2& Feature invariance & Summation of a feature is 0, and\\
&& standard deviation of a feature is 0\\
&& at bot and normal user, respectively\\

\hline
\end{tabular}%
\end{table}

\begin{table*}[t]
\caption{List of essential features}\label{tab:essential_feature_list}
\centering

\begin{tabular}{|c|l|l|l|}
\hline
No. & Type & Feature & Description\\

\hline
1 & Item & Number of Items & Total number of items a character owns\\

\hline
2 & Cash & Total Cash & Total amount of cash a character owns\\
3 & & Cash in Account & Amount of cash a character carries in inventory\\
4 & & Cash in Character Bank & Amount of cash a character stores in warehouse\\
5 & & Cash in Vendor & Amount of operating cash handled by transaction vendors\\
& & & such as sales agent or auction house\\

\hline
6 & Evaluated & Evaluated Asset Value & Sum of monetary value of cash and all items\\
& Asset Value& & evaluated by default price\\

7 & & Mailing Asset Value & Sum of monetary value of all items in sent and received by mail\\
& & & evaluated by default price\\

8 & & Evaluated Asset value & Sum of monetary value of cash and all items\\
& & in character bank & stored in character's warehouse evaluated by default price\\

9 & & Evaluated Asset & Sum of cash and monetary value of all items\\
& & in account bank & stored in account's warehouse evaluated by default price\\

\hline
\end{tabular}%
\end{table*}

\begin{figure*}[b] 
\centering
\includegraphics[width=14cm]{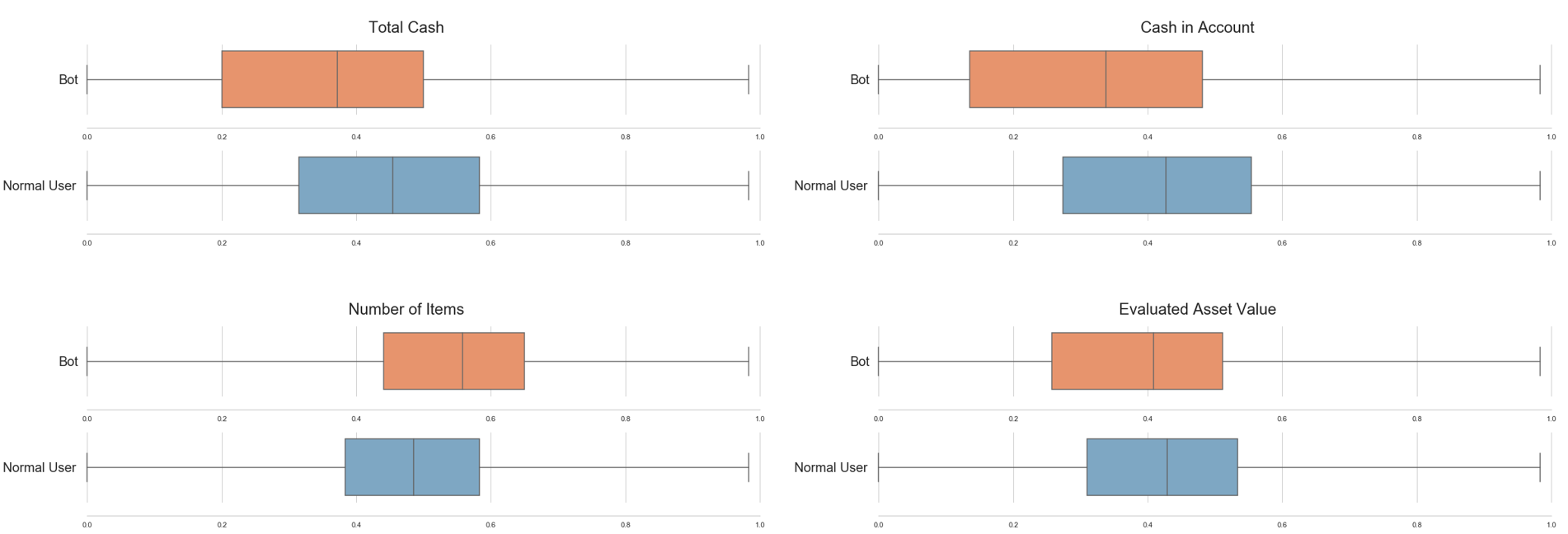}
\caption{Chosen boxplots of scaled feature distribution} 
\label{fig:feature_distribution}
\end{figure*}

\subsection{Feature Engineering}
Feature engineering process takes significant position in model performance. To make log-level raw data into trainable form, we managed two steps of feature engineering: eliminating non-influential features, sliding time window with scaling.

\subsubsection{Eliminating non-influential features}
To scrutinize change of financial data, we eliminated features without meaningful level of change. To leverage deep neural networks on bot detection, model should learn different dynamics of data between bot and normal user. If a feature with similar pattern is provided to the model, it blurs weights and bias of neural network. Thus, we set rules described in Table~\ref{tab:rules} to filter non-influential features.

If a feature is caught in Rule 1, it implicates model cannot learn any different pattern. A feature with same value between bot and normal user does not make any different pattern, thus we dropped features fulfilling Rule 1. Rule 2 proves that feature value is 0, creating sparsity at the training data. If we map training data into vector space, a feature filled with zero value creates sparse input vector. As sparse input vector blurs computation of model parameters, we dropped features fulfilling Rule 2 to reduce sparsity. After applying two rules above, we extracted essential features shown in Table~\ref{tab:essential_feature_list} from log-level raw data.

\subsubsection{Sliding time window with Scaling}
Log-level raw data necessitates transformation process to be in trainable form. As deep neural networks effectively learn from fixed length of data, long time-series data require cutting process. We set fixed length of time window, and generated training data by sliding along log-level raw data. Moreover, we applied scaling process to mitigate different scales of data. By calculating each feature with the equation below, we transformed data to be scaled between 0 and 1. Lastly, we labeled transformed data same as its original data.
\[ X_{normalized} = \frac{X_i - Min(X_i)}{Max(X_i) - Min(X_i)}\]\

We compared statistical difference of features to check whether they are distinct along with labels. As illustrated in Figure~\ref{fig:feature_distribution}, features such as total cash, cash in account, and number of items show different distribution at the same scale. Considering these differences, we clarified financial features can differentiate bots from normal users.

\subsection{Modeling}
We employed RNN as detection algorithm. RNN is one of deep neural networks that past computation result influences next computation of model parameters. Previous research on neural networks have shown RNN explores well with time-series data \cite{connor1991recurrent}. We analyzed financial data entangles temporal dynamics, thus utilized recurrent architecture to model. Among various types of RNN, we set LSTM neural networks considering its performance. We provided training data into LSTM network to identify bots and normal users. Training was performed along prefixed size of batches, and batch normalization was applied for better performance. For stabilized training, we also applied regularization techniques to evade the model's overfitting problem. After the training process finished, we validated the model with test dataset, and record its performance with four metrics: accuracy, precision, recall, and F1-score.

\section{Experiment Result}

To gain enough records of training, we separated monthly dataset into weekly basis. We randomly mixed data, and performed 10-fold cross validation for assured result. Table~\ref{tab:experiment_result} shows the result of experiment with evaluation metrics.

\begin{table}[h!]
\caption{Bot detection experiment result}
\label{tab:experiment_result}
\centering
\begin{tabular}{|c|c|c|c|c|}
\hline
Experiment & Accuracy & Precision & Recall & F1 Score\\
\hline
Week 1 & 0.9494 & 0.9385 & 0.9759 & 0.9490\\
Week 2 & 0.9401 & 0.9168 & 0.9831 & 0.9488\\
Week 3 & 0.9487 & 0.9237 & 0.9886 & 0.9551\\
Week 4 & 0.9509 & 0.9103 & 0.9861 & 0.9476\\
\hline
Average & 0.9473 & 0.9223 & 0.9834 & 0.9501\\
\hline
\end{tabular}
\end{table}

We interpret trained model detects bots with reasonable performance. Experiment result shows relatively lower precision than recall, implicating existence of false positive error. Considering characteristics of bot detection, we evaluate this error is allowable. One of primary goals of bot detection is making a list of doubtful accounts rather than blocking accounts. If game company blocks normal user showing similar pattern as bots, it directly generates dissatisfaction or even annoy of user. Thus, bot detection prioritizes figuring out doubtful characters rather than direct blocking. In this viewpoint, we evaluate our model efficiently detects bots only with a few financial features. We expect nsembling our model with other methods would achieve more precise detection, which is further illustrated in following Discussion section.

\section{Discussion}
\subsubsection{Ensembling for precise detection}
Ensembling our model with other methods would build powerful bot detection system. To hedge risk of false positive error, bots are not decided by a single detection method. To reduce false positive errors, even human-based detection method in the past required many skilled observers. In similar way, we believe more detection methods are required to collaborate for industry-level application. For features of ensembling, we would leverage other necessary conditions of game bots. Transaction with other characters or Non-Player Characters (NPC) can be utilized as features. Considering more information of bots, detection model would be improved resulting powerful performance.

\subsubsection{Applicability into other MMORPGs}
On following research, we would apply this model in other MMORPGs. With the growth of game industry, game companies operate multiple games at the same time. To put in game company's shoes, running different detection methods for different games is cost-ineffective. If detection method is applicable in multiple games, it creates economies of scale. We interpret features used in our model are commonly found. Previous research on bot network also suggested financial features are common traces of modern MMORPGs \cite{kwon2016crime}. Therefore, we expect our model is applicable in other games to be used in industry-level practices.

\section{Conclusion}
In this work, we proposed employment of financial analysis to game bot detection in real MMORPG ecosystem. Previous studies showed efficient bot detection methods, but underlying problem was variance of bot's activity. If bot developers change activity or make them act like normal users, it becomes resource-consumable to update detection methods. To overcome this limitation, we utilized financial information of characters which are necessary conditions of bot's activity. Considering purpose of bots related to RMT, patterned change of financial status is inevitable for bot characters. Through experiments, our results reveal that financial information are highly related to bot's activity, suggesting effectiveness at bot detection. Leveraging time-series dynamics of financial data, our LSTM neural networks efficiently captured pattern of bots. We validated this efficiency with actual game data of Aion, a famous MMORPG developed by NCSOFT.

Our detection model establishes essential contributions. First and foremost, our model establishes sustainable detection method as bots are hard to evade patterned financial changes. Moreover, our model is applicable into other games as utilized features are common in modern MMORPG. Unlike macroscopic analysis of bot network, LSTM neural network is eligible to detect individual characters to generate list of doubtful characters. Finally, our model enhances security of detection as LSTM neural network is a black-box model making bot developers hard to predict detection thresholds. There still exists a room for improvement. With ensembling our model with other necessary conditions of game bot, we would like to achieve qualified bot detection system with powerful performance. To assure generality of our model, we plan to check model's performance on other actual MMORPG data. In pursuit of sustainable, and general bot detection, we will further research on financial analysis in MMORPG ecosystem.

\section*{Acknowledgements}
This work was supported under the framework of international cooperation program managed by National Research Foundation of Korea(No.2017K1A3A1A17092614). 

\bibliographystyle{IEEEtran}
\bibliography{reference.bib}

\end{document}